\documentclass{llncs}
\usepackage{graphicx}
\usepackage{graphicx}
\usepackage{amsmath}
\usepackage{cite}
\usepackage{enumerate}
\usepackage{nicefrac}
\usepackage{color}

\usepackage[bottom]{footmisc}

\begin{document}
\title{Enhancing an eco-driving gamification platform through wearable and vehicle sensor data integration}
%
%
\author{Christos Tselios\inst{1} \and 
Stavros Nousias\inst{1} \and Dimitris Bitzas\inst{1} \and Dimitrios Amaxilatis\inst{2} \and Orestis Akrivopoulos\inst{2} \and Aris S. Lalos \inst{1,4} \and Konstantinos Moustakas\inst{1} \and Ioannis Chatzigiannakis\inst{3} 
}
\authorrunning{C. Tselios et al.}
%
\institute{University of Patras, Achaia 26500, Greece \\ 
\email{\{tselios, nousias, bintzas, aris.lalos, moustakas\} @ece.upatras.gr}
\and
SparkWorks ITC Ltd, Alticham, United Kingdom \\
\email{\{d.amaxilatis, akribopo\} @sparkworks.net}\\ 
\and Sapienza University of Rome, Italy \\
\email{ichatz @diag.uniromal.it}
\and
Industrial Systems Institute, ATHENA Research Center, Greece }
\maketitle              
\begin{abstract}
As road transportation has been identified as a major contributor of environmental pollution, motivating individuals to adopt a more eco-friendly driving style could have a substantial ecological as well as financial benefit. With gamification being an effective tool towards guiding targeted behavioural changes, the development of realistic frameworks delivering a high end user experience, becomes a topic of active research. This paper presents a series of enhancements introduced to an eco-driving gamification platform by the integration of additional wearable and vehicle-oriented sensing data sources, leading to a much more realistic evaluation of the context of a driving session.  

\keywords{Gamification  \and Eco-driving \and Sensors.}
\end{abstract}
\section{Introduction}
\label{sec:intro}
Road {\let\thefootnote\relax\footnote{Part of this work has been supported by H2020-ICT-24-2016 Project GamECAR (Grant No. 732068) and H2020-SC1-DTH-2018-1 Project SmartWork (Grant No. 826343)}} transportation greatly aggravates environmental pollution given the fact that approximately 30\% of the total CO2 emissions worldwide derive from internal combustion engines used in motorized vehicles \cite{SANTOS201771}. Abolishing highly polluting engines and replace every car with an electric one is probably the proper method for tackling pollution, however this will not happen overnight. It is therefore essential to motivate drivers towards adopting a more eco-friendly driving style for actually having a chance of achieving any substantial ecological or financial benefit. Alas, an individual's driving style is fused by a series of factors often including psychological, cultural and societal ones which are difficult to identify as well as change. Yet, gamification proved to be an effective tool towards guiding targeted behavioural changes and focusing on such methods could be something worth considering. 

Gamification is directly correlated with the introduction of specialized, game-related guidelines utilized to facilitate the subconscious adoption of a desired behaviour. Users are encouraged to follow specific ways of dealing with issues through the stimulation of intrinsic personal traits, often referred as \textit{core drives} \cite{chou2019actionable}. The stimulation of core drives such as (i) accomplishment, (ii) empowerment of creativity, (iii) ownership, (iv) relatedness, (vi) scarcity, (v) unpredictability and (vi) curiosity, greatly increase user interest, commitment and involvement towards concluding any given task. Individuals are motivated and essentially guided to adopt pre-defined execution patterns via a playful experience, specifically designed to elevate awareness and negative outcome provision ability, all through participation in a seemingly simplistic game.  

As stated in \cite{chou2019actionable} the main pillars of gamification combine a large variety of verticals such as technological platforms, user experience and game mechanics optimization, motivational psychology and behavioural economics, all glued together through an underlying business logic. 
Following a more person-oriented approach, Kim et. al \cite{kim2011gamification} suggested differentiating between intrinsic and extrinsic rewards 
while Yu-Kai Chou proposed the \textit{Octalysis} actionable gamification framework \cite{chou2019actionable} which incorporates previous approaches in an attempt to realistically determine elements which make games both meaningful and fun. 

It becomes obvious that eco-driving could be significantly improved through a gamification platform which will guide users towards adopting a more environmental friendly driving attitude. The rest of the paper is organized as follows: Section 2 presents some of the elements for designing realistic eco-driving gamification platform while Section 3 focuses on how accumulating data from vehicle and wearable sensors boosts the platform's realism and further engages users. Section 4 describes the prototype implementation process and finally, Section 5 draws conclusions and summarizes the paper.

\section{Gamification framework prototype for improving eco-driving behaviour}
\label{sec:gamification}
The original prototype of the gamification platform was developed primarily to motivate users to adopt an eco-friendly driving style through real-time push notifications and auto-generated feedback. Safe driving was not to be compromised in any way therefore all platform-oriented messages were designed to be non-intrusive allowing drivers to focus on the road. The \textit{Octalysis} actionable gamification framework was taken into consideration, yet without neglecting complementary gamification methodologies, each with significant benefits. All playful interventions were adapted to the characteristics of the driver by a dedicated module that was based on cutting-edge user models for expressing one's driving behaviour.     

\begin{figure}[t!]
\centering
\includegraphics[width=0.9\textwidth]{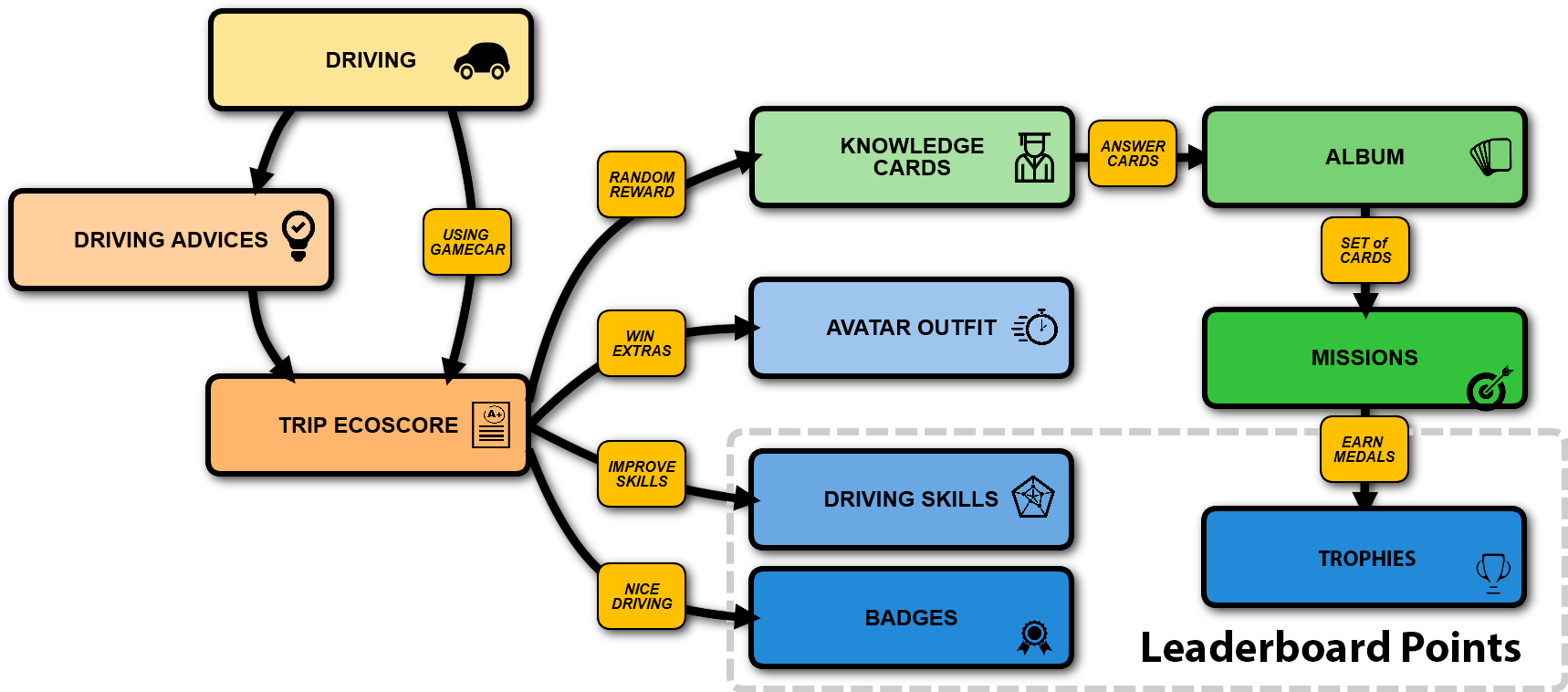}
\caption{Gamification flowchart designed to improve eco-driving behaviour}
\label{fig:design}
\end{figure}

The interaction between users and the gamification platform was designed to be made via their real driving sessions, the tracking of which was rendered possible by dedicated equipment carried on board. The cornerstone of this equipment was the driver's smartphone and the variety of features it integrates, such as the accelerometer and the GPS Navigation system. Additional, more direct parameters related to one's specific driving skills such as hard/soft braking as well as an overall aggressiveness score linearly linked to the total number of braking/accelerating actions per minute, were calculated, associated with the driver's profile and stored for future reference. Figure \ref{fig:design} presents a flowchart of the overall process which allowed drivers to improve their eco-driving skills through serious gaming \cite{NOUSIAS2019103}. All elements are tailor-made to automatically align with the various levels of player experience which progressively increases as the game progresses, while each level has distinctive characteristics, features and objectives. 

As shown in Figure \ref{fig:design}, each driving session mainly contributes to the creation of the \textit{Trip Ecoscore} which is the most important gamification mechanism. It provides information to the user regarding the eco-driving quality of the previous session and is immediately correlated to the overall progress. Moreover, it acts as the basis for every possible reward generation, thus is directly linked with the core drives of the \textit{Octalysis} gamification framework \cite{chou2019actionable}. Obtaining an acceptable Trip Ecoscore is the key for (i) earning \textit{Badges}, (ii) collecting points which improve the overall \textit{Driving Skills}, (iii) be rewarded with access to \textit{Knowledge Cards} which unlock \textit{Missions} and consequently \textit{Trophies}. Badges, Driving Skill score and Trophies are visible to the platform's \textit{Leaderboard}, an online \textit{Hall of Fame} where eco-driving record scores are kept. In addition, each user owns an \textit{Avatar} acting as their virtual representation, starting from a basic avatar and gradually unlocking multiple avatar outfit parts during their progression in the game. The avatar is displayed in the driver's \textit{Profile}, a crucial gamification element as it provides a clear view of a player's progression in all aspects of the game. 


\section{Increasing realism through sensor data assembly}
\label{sec:prototype}
\begin{figure}
\centering
\includegraphics[width=0.75\textwidth]{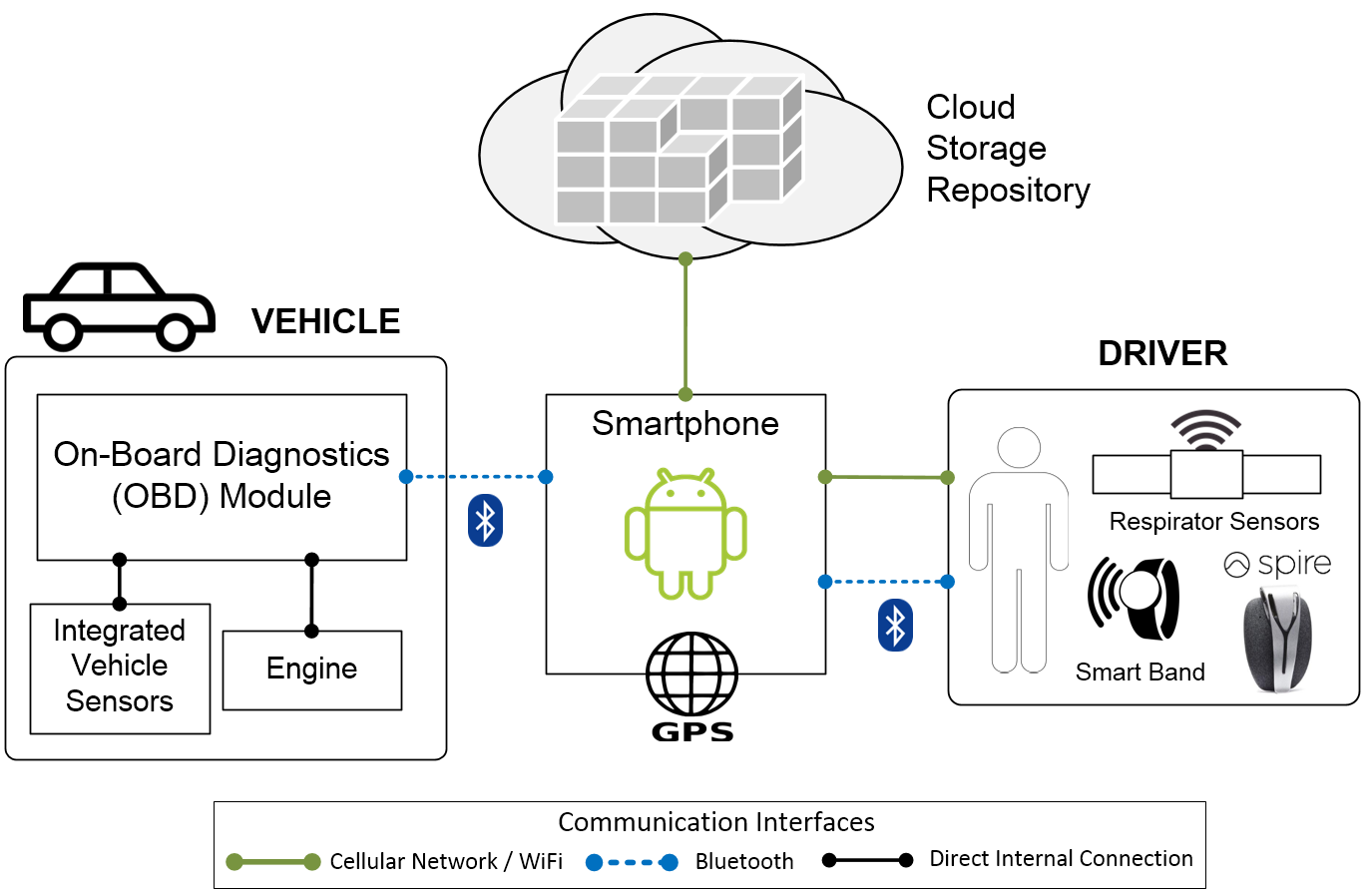}
\caption{High level architectural diagram of the Mobility Monitoring Network prototype } \label{architecture}
\end{figure}

Despite the overall functionality of the implemented prototype, some issues were soon identified. Relying on the smartphone's accelerometer for getting a rough estimation of the vehicle's driving condition proved to be an inaccurate approach. The system failed to expose the significant variations in cruising speed when those occurred after smooth acceleration or braking, thus returned an erroneous Trip Ecoscore value estimation. Fast moving vehicles were treated the same way with slow moving ones, when speed remained consistent for a longer duration of time, a behaviour which does not comply with the real consumption and CO2 emission. This situation was tackled with the introduction of additional metrics obtained from two additional categories of sensing devices each contributing with specific bits of information for filling the gaps and reproduce the context of each route in a realistic manner. A high level architecture of the new Mobility Monitoring Network prototype is presented in Figure \ref{architecture}. 

For implementing a driving behaviour monitoring framework which will reveal the condition of both the vehicle as well as the driver, it is essential to record a series of physiological, behavioural, environmental and vehicle parameters. Therefore, input was obtained from two major sensing categories, vehicle and driver-oriented.  
\begin{enumerate}
    \item \textbf{Vehicle sensors} measuring gear shifts, tire pressure, temperature and oil are already integrated to modern cars by virtually every manufacturer, linked through an internal controller area network (CAN) for ensuring robust communication. Data can be retrieved using an on-board diagnostics (OBD) controller and reveal a detailed log of the vehicle's condition on any given time. 
    \item \textbf{Wearables} are attached to the driver's body and non-intrusively record aspects of his physiological condition. The obtained traces (raw data) are encrypted and stored in the internal memory of the device remaining available for further process.
\end{enumerate}

The smartphone retained the role of the main coordination node due to the large number of communication interfaces, caching and processing capability it supports as well as the auxiliary data sources it incorporates. The overall  functionality of establishing communication with all necessary devices, retrieve data and process them accordingly was integrated into the gamification application of the original prototype. Moreover, techniques for data-preprocessing, customized for extracting the most significant information \cite{nousiasl2018managing} were also integrated. For the preliminary prototype evaluation and without the loss of generality, only data regarding vehicle speed, engine rounds-per-minute (RPM), braking, throttle position and driver heartbeats per minute were collected for creating matrices stored in a per-trip fashion. The smartphone acted as a data aggregator that accumulated sensor values, added a timestamp and created a .CSV file. This file was also populated by additional content retrieved from the smartphone's GPS which indicated the exact positioning of the driver/vehicle. At the end of each route, the .CSV file was transmitted to the Cloud Storage Repository using affiliated WiFi or 4G/LTE network connection.

\section{Gamification accuracy}
\label{sec:implementation}
Unlike the original prototype which only relied on acceleration and speed changes, it is now possible to obtain additional and highly accurate information for braking, RPM and throttle position. This lead to a more precise eco-score function which penalizes high level of engine RPM, is associated with gear shift-up while cruising and subtracts points for cases of abrupt braking and high acceleration, in correlation with the driver's heartbeat condition. Moreover, a complementary metric, the aggressiveness score penalizes high lateral acceleration, abrupt braking and high variances in throttle position and RPM.  Both metrics contribute to the Trip Ecoscore value which remains crucial for every stage of the game. Eco-score is comprised of five parameters, shift-up parameter, braking parameter, acceleration parameter, RPM parameter and cruising parameter. All parameters are computed within a 30 seconds temporal window. For each metric, we utilize a sigmoid function $\sigma(x)$ and a weighted histogram function that assigns each event to a bin depending on the intensity of the event. $\sigma(x)$ is defined as
\begin{equation}
    \sigma(x)=a_2+\left(\frac{a_1}{a_4+e^{\left(a_3 \cdot [x-x_0] \right)}}\right)
\end{equation}
where parameters $a_{1}$ to $a_{4}$ are experimentally defined.

The aggressiveness score calculation is based on parameters such as (i) engine RPM variance, (ii) braking intensity and (iii) acceleration magnitude. As expected, high variance of engine RPM, especially when paired with higher than normal heartbeat rate, indicates driver nervousness and improper vehicle handling. In most cases, significant fluctuations in engine RPM occur when drivers are not focused on smooth cruising and overall safe driving but tend to constantly accelerate, thus minimizing their fuel efficiency, eco-friendliness and consequently the overall road safety. Given a temporal window of $t=30$ seconds, the aggressiveness score $AG_{RPM}$ derived only from engine RPM fluctuation is
\begin{equation}
    AG_{RPM}=\frac{s^2_{RPM}}{\mu}
\end{equation}
where $s^2_{RPM}$ is the engine RPM variance and $\mu$ is an experimentally defined parameter.
Another metric not available in the first prototype is Braking Intensity $BI$, calculated by correlating vehicle deceleration and braking duration, during the online analysis phase. This metric cumulatively contributes to the computation of aggressiveness score given the fact that abrupt braking events indicate both poor planning and low level of situation awareness regarding road conditions from the driver's side. The braking intensity aggressiveness score $BI_{AG}$ is equal to 
\begin{equation}
    BI_{AG}=\frac{B_{a}}{B_{a}+B_{s}}
\end{equation}

where $B_a$ is the number of abrupt braking events and $B_s$ is the number of smooth braking events within the aforementioned temporal window. 


\section{Conclusions}
\label{sec:conclusions}
This work presented an ongoing effort to improve a fully functional eco-driving gamification platform prototype by integrating additional datasets obtained by vehicle and wearable sensors. The introduction of a holistic and highly accurate data retrieval platform greatly optimizes the overall in-game experience, which is now directly linked and fully correlated with the actual driver, vehicle and route conditions. Preliminary results indicate that drivers were able remain actively engaged in all gaming activity, while the ultimate goal of endorsing a more eco-driving behaviour is achieved.

%
%
%
\bibliographystyle{splncs04}
\bibliography{samplepaper}

\begin{thebibliography}{1}
\providecommand{\url}[1]{\texttt{#1}}
\providecommand{\urlprefix}{URL }
\providecommand{\doi}[1]{https://doi.org/#1}

\bibitem{chou2019actionable}
Chou, Y.k.: Actionable gamification: Beyond points, badges, and leaderboards.
  Packt Publishing Ltd (2019)

\bibitem{kim2011gamification}
Kim, A.J.: Gamification 101: Design the player journey. Retrieved August
  \textbf{5}, ~2014 (2011)

\bibitem{NOUSIAS2019103}
Nousias, S., Tselios, C., Bitzas, D., Amaxilatis, D., Montesa, J., Lalos, A.S.,
  Moustakas, K., Chatzigiannakis, I.: Exploiting gamification to improve
  eco-driving behaviour: The gamecar approach. Electronic Notes in Theoretical
  Computer Science  \textbf{343},  103 -- 116 (2019).
  \doi{https://doi.org/10.1016/j.entcs.2019.04.013},
  \url{http://www.sciencedirect.com/science/article/pii/S1571066119300167}, the
  proceedings of AmI, the 2018 European Conference on Ambient Intelligence.

\bibitem{nousiasl2018managing}
Nousiasl, S., Tseliosl, C., Orfila, O., Jamson, S., Mejuto, P., Amaxilatis, D.,
  Akrivopoulos, O., Chatzigiannakis, I., Lalosl, A.S., Moustakasl, K., et~al.:
  Managing nonuniformities and uncertainties in vehicle-oriented sensor data
  over next generation networks. In: 2018 IEEE International Conference on
  Pervasive Computing and Communications Workshops (PerCom Workshops). pp.
  272--277. IEEE (2018)

\bibitem{SANTOS201771}
Santos, G.: Road transport and co2 emissions: What are the challenges?
  Transport Policy  \textbf{59},  71 -- 74 (2017).
  \doi{https://doi.org/10.1016/j.tranpol.2017.06.007},
  \url{http://www.sciencedirect.com/science/article/pii/S0967070X17304262}

\end{thebibliography}
\end{document}